\documentclass[aps,preprint,preprint]{revtex4}
\usepackage{epsfig}
\usepackage{amsmath}
\usepackage{epsfig}
\usepackage{amssymb}

\begin{document}

\title
{\bf Comment on ``Quantization of the damped harmonic oscillator'' [Serhan et al, J. Math. Phys. {\bf 59}, 082105 (2018)]} 
\author{Zafar Ahmed$^{1,*}$, Sachin Kumar$^2$, and Abhijit Baishya$^3$}
\affiliation{$~^1$Nuclear Physics Division,  Bhabha Atomic Research Centre, Mumbai 400 085, India\\
$~^*$Homi Bhabha National Institute, Mumbai 400 094 , India	\\
	 $~^2$Theoretical Physics Section, Bhabha Atomic Research Centre, Mumbai 400 085, India
	\\
	$~^3$Human Resource Development Division, Bhabha Atomic Research Centre, Mumbai 400 085, India}
\email{1:zahmed@barc.gov.in, 2: sachinv@barc.gov.in,  3: abhijitb498@gmail.com,}   
\date{\today}

\begin{abstract}
A recent paper [J. Math. Phys. {\bf 59}, 082105 (2018)] constructs a Hamiltonian for the (dissipative) damped harmonic oscillator. We point out that non-Hermiticity of this  Hamiltonian has been ignored to find real discrete eigenvalues which are actually non-real. We emphasize that non-Hermiticity in Hamiltonian is crucial and it is a quantal  signature of dissipation.
\end{abstract}
\maketitle
Heuristically, by making a Hamiltonian non-Hermitian one can account for dissipation of energy and matter. Presently interesting frameworks are being developed [1] wherein classical dissipative systems can be associated with non-Hermitian quantum Hamiltonians. The simplest model of dissipative systems is classical damped harmonic oscillator (DHO) whose equation of motion is  
\begin{equation}
\ddot q + \lambda \dot q +\omega^2 q=0.
\end{equation}
Recently, this has been transformed to a Hamiltonian $H$ [2]. 
\begin{equation}
H=\frac{p_y^2}{2m}+ \frac{1}{2} m \omega^2 y^2 +\frac{\lambda y p_y}{2}.
\end{equation}
We point out that the non-Hermiticity of $H$ due to $(y p_y)^\dagger= yp_y-i\hbar$ has been ignored and  real discrete eigenvalues 
\begin{equation}
E_n=(n+1/2)\hbar \sqrt{\omega^2-\lambda^2/4}
\end{equation}
have been obtained [2] by some N-U method [2]. Using $yp_y-p_y y=i\hbar$, $H$ can be re-written as
\begin{equation}
H=\frac{p_y^2}{2m}+ \frac{1}{2} m \omega^2 y^2 +\frac{\lambda (y p_y + p_y y)}{4} + \frac{i\hbar \lambda}{4}.
\end{equation}
This can be re-written as
\begin{equation}
H=\frac{(p_y+ m\lambda y/2)^2}{2m}+\frac{1}{2} m(\omega^2-\lambda^2/4) y^2 + \frac{i\hbar \lambda}{4}.
\end{equation}
Consider the eigenvalue equation: $H\psi=E\psi \Rightarrow \eta H \eta ^{-1} \eta \psi = E \eta \psi$ (see [3]). Let us choose $\eta= \exp(im\lambda y^2/4\hbar)$. We find that 
\begin{eqnarray}
e^{(im\lambda y^2/4\hbar)}~[p_y+m\lambda y/2] ~ e^{-(im\lambda y^2/4\hbar)}= p_y,\nonumber \\
e^{(im\lambda y^2/4\hbar)}~[p_y+m\lambda y/2]^n ~ e^{-(im\lambda y^2/4\hbar)}= p_y^n.
\end{eqnarray}
Consequently,
\begin{equation}
\eta H \eta^{-1} =\frac{p_y^2}{2m}+\frac{1}{2} m(\omega^2-\lambda^2/4) y^2 + \frac{i\hbar \lambda}{4}.
\end{equation}
Finally the quantization of (7) leads to 
\begin{equation}
E_n=(n+1/2)\hbar \sqrt{\omega^2-\lambda^2/4} +\frac{i\hbar \lambda}{4},
\end{equation}
which are complex and it is rightly so, as the Hamiltonian (2) is non-Hermitian. One needs to highlight  that the Hamiltonian (2) for the damped harmonic oscillator (1) is crucially non-Hermitian, this will help in the development of  quantization of dissipative systems. 
\section*{References}
\begin{enumerate}
\item Eva-Maria Graefe, Michael Honing, and Hans Jurgen Korsch, J. Phys. A: Math. Theor. {\bf 43} 075306 (2010).
\item M. Serhan, M. Abusini, Ahmed Al-Jamel, H. El-Nasser, and Eqab M. Rabei, J. Math. Phys. J. Math. Phys. {\bf 59}, 082105 (2018).
\item Z. Ahmed, Phys. Lett. A {\bf 294} 287 (2002).
\end{enumerate}
\end{document}